# Computational Solutions for Today's Navy

*New methods are being employed to meet the Navy's changing software-development environment.*

▶ Frank W. Bentrem, Ph.D., John T. Sample, Ph.D., and Michael M. Harris

The Naval Research Laboratory (NRL) is the corporate laboratory for the United States Navy. Part of the mission of NRL is to provide "broadly based applied research and advanced technology development programs in response to identified and anticipated Navy and Marine Corps needs". Advancing the state-of-the-art in scientific computation is important in both applied research and technology development. In this article, we relate current computational trends in naval technologies as well as the trends we anticipate over the next few years. We also present solutions for some of the current and coming challenges we face.

Naval operations can benefit greatly from a highly accurate and timely knowledge of the environment. Specifically, mine-hunting operations require knowledge of the seafloor composition and oceanographic conditions which include the temperature, density, water clarity, and depth of the water, the speed of ocean currents, and wave heights. Naval personnel use a combination of historical, real-time, and predicted data about these conditions.

### Sea-based computing

*Sea basing* refers to a major thrust in future naval capabilities from the Office of Naval Research where "logistics, shipping, and at-sea transfer technologies support joint operational independence". This applies equally well to environmental data processing. Indeed, in recent years, we have noticed an increasing shift towards computationally intensive data processing in the field, fleet, and during survey operations as opposed to collecting and processing data at centralized facilities. Contributing to this shift is the development and manufacturing of affordable and compact (yet powerful) workstations. Since computers meeting both budget and space constraints are now widely available, these machines may be installed in the ship or other vehicle along with the necessary software. In this way, rather than depending on the reliability and data-transfer rates of the communication links, operators can use onboard data processing to make real-time de-

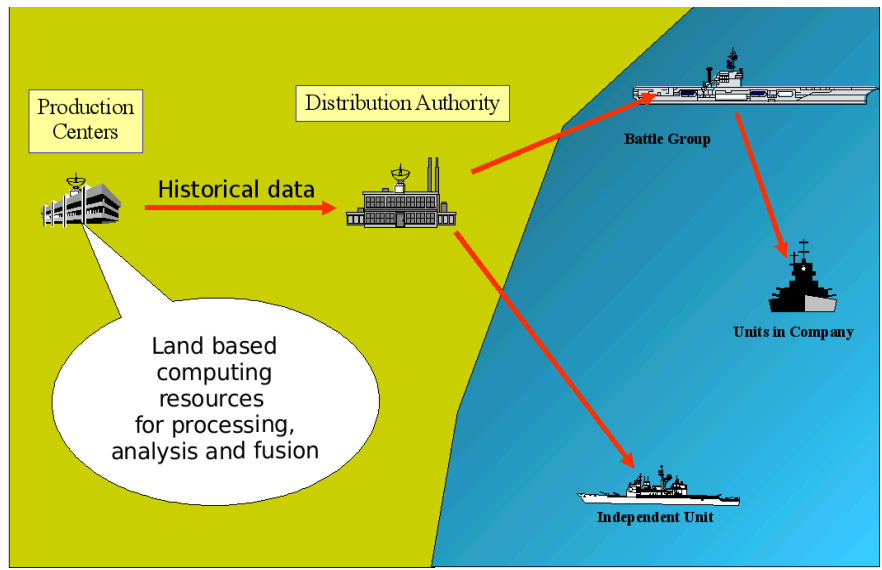

**OLD WAY** of doing business: Computing power is located at shore-based production facilities that determine and distribute the "best" environment.

cisions. While not exactly a new method, defense organizations are increasingly processing data in the field or at sea.

The most important demands on computing systems resulting from the sea-based computing model are twofold--*portability* and *efficiency*. While software development targets usability in the field, expert analysts must also be able to use the same software in their office spaces, either for further analysis or to process archived data, often with a different platform. Cross-platform computer languages such as C++ with Qt (Trolltech) or Java (Sun Microsystems) are especially useful for these applications. Not only must computing solutions be portable, processing in the field or at sea requires large amounts of data to be analyzed efficiently, in fact, in real-time. Modern compiled languages (e.g. C, C++, Fortran 95) are well suited for speedy computations, however, good *software design* is the most important factor in program efficiency.

### Through-the-Sensor technology

In addition to sea-based computing, there is a definite trend towards techniques that allow environmental data to be extracted from *tactical fleet sensors*[1]. This requires computer systems to have the flexibility to acquire and process data from combat sensors rather than sensors designed for scientific purposes.

Software aids are used to help make tactical decisions by military operators. These decision aids use environmental inputs, and then output optimal methods for finding targets. Actual environmental (not historical) data is needed to yield accurate answers.

Through-the-Sensor (TTS) techniques use tactical sensors to acquire environmental information in situ. This new information is used to refresh historical holdings. Using TTS data, decision aids can provide answers based on actual versus historical conditions. The challenge is to collect, process, fuse, and disseminate TTS data in near real-time. The benefit is lowered operational risk

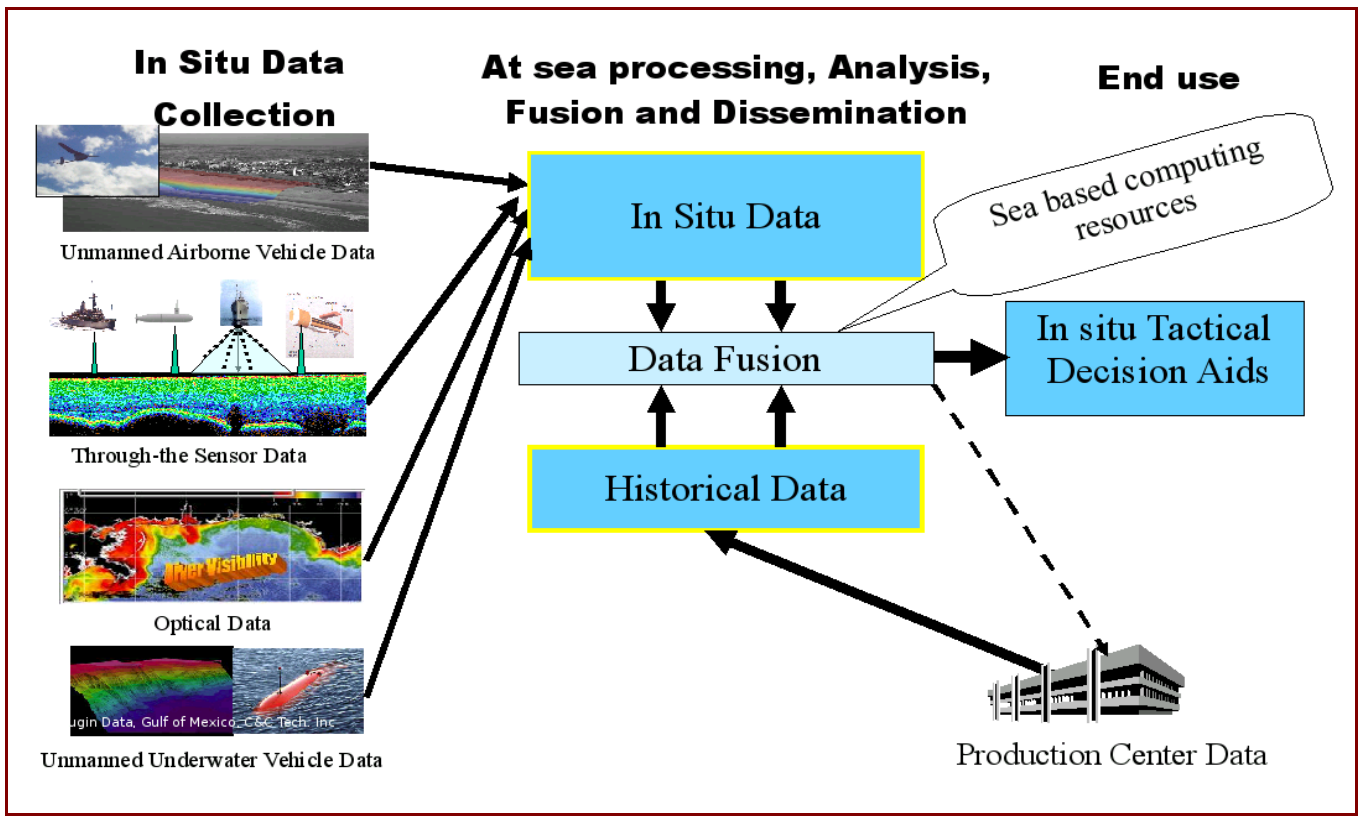

**NEW WAY** of doing business: Computing power is located at sea and ashore. "Best" environment is determined on-scene using new *in situ* data to refresh historical data.

and accurate time lines based on actual conditions.

### Recent technological solutions

NRL and the Naval Oceanographic Office have recently developed the Environmental Post Mission Analysis (EPMA) system to manage the integration of these types of data. The EPMA collects data from tactical sidescan sonars (Through-the-Sensor technology), supercomputer generated numerical models, and historical/climatological databases. It uses a variety of fusion algorithms to merge those sources into a single best view of the naval environment. The EPMA is developed in C++ with Qt and is used on Windows, Linux, and UNIX platforms; all of which appear both at various Navy installations and on Naval ships (sea-based technology).

Another recent example of the need for sea-based technology is the requirement for sediment identification at the ocean floor during naval oceanographic surveys. In the past, locations for numerous sediment core samples were chosen prior to embarking on a survey. NRL has since developed SediMap®[2,3] to enable the mapping of sediments during the survey itself. SediMap® uses the Java Runtime Environment for the capability of executing on a number of different platforms. To maximize efficiency, a number of tests were conducted to determine the optimal convergence rate for sediment-size estimates without sacrificing accuracy. The software runs in real-time as data is collected and may either be installed on the onboard computer or simply run off the CD.

Further examples of Through-the-Sensor data acquisition are the extraction of bathymetry and surficial seafloor sediment types from the UQN-4 fathometer on mine counter-measures (MCM) ships, the BQN-17 submarine fathometer, and the AN/AQS-20A mine-hunting system (which also yields multibeam bathymetry). These data are needed in anti-submarine and mine-hunting decision aids to determine tactics.

### Conclusions

To better utilize large quantities of environmental data, the Navy is moving towards *sea-based computing*, which allows decisions to be made in real-time at sea. Also, *Through-the-Sensor technology* is being developed for existing/future fleet combat sensors to be used in tactical decision aids. These advances will help the US Navy maintain sea supremacy in today's fast-paced war-fighting environment.

*Frank W. Bentrem, Ph.D. is a research physicist, John T. Sample, Ph.D. is a computer scientist, and Michael M. Harris is the branch head for Mapping, Charting, and Geodesy, all three authors with the Marine Geosciences Division at the Naval Research Laboratory. They may be reached at frank.bentrem@nrlssc.navy.mil.*